\input epsf
\documentstyle[prc,aps,twocolumn]{revtex}
\begin{document}
\leftmargin=-1cm
\title{Calculation of the number of partitions with constraints on the fragment 
size}
\author{P. D\'esesquelles\\
Institut de Physique Nucl\'eaire, Bt. 100, 15 rue Georges
Cl\'emenceau,\\
F91406 Orsay Cedex, France}

\maketitle

\begin{abstract}
This article introduces recursive relations allowing the calculation of the
number of partitions with constraints on the minimum and/or on the maximum
fragment size.
\end{abstract}

\section {Introduction}

A partition is an ensemble of positive integers with a given sum\footnote{For
example, the integer 3 has 3 partitions : \{3\}, \{2,1\} and \{1,1,1\}.}. A
partition can also be seen as a way to break a piece of discret matter into
{\em fragments}. The number of partitions of a given integer is a quantity
which is useful in various fields. In the so-called minimal information model
\cite{Bib_Eng84_Eng91_Bot00} for example, all the partitions have the same
probability\footnote{In the following, the total number
of partitions of the integer $S$ will be noted $N(S)$, a partition will be
noted {\bf n}$:(n_1,\ldots,n_S)$, $n_s$ being the number of integers $s$ in the
partition.} $P({\bf n}) = 1/N(S)$. This result is obtained by application of the minimal information
principle (or maximum entropy), information being defined as $\sum_{{\bf n}}
P({\bf n}) \ln{P({\bf n})}$. When this equation is differentiated under the
only constraint $\sum_s s\,n_s = S$, all probabilities are found to be equal
\cite{Bib_Aic84}.

\subsection{Physical Interest}

In many cases, it is interesting to distinguish one or more classes of
fragments according to their size. In percolation, for instance, the
sub-critical events are defined by the fact that they contain one particular
fragment which is refered to as {\it infinite} or {\it percolative} in the
sense that it connects the two extremes of the lattice \cite{Bib_Sta85_Cam88}.
In the case of conductive bonds, it allows an electric current to circulate
between two electrodes placed on opposite surfaces of the lattice. In the case
of a coffee-machine, the percolative cluster defines a path which allows the
vapor to traverse the grounds. In Nuclear Physics, other classes of fragments
are distinguished. The {\it intermediate mass fragments} ({\sc imf}) are ions
resulting from the violent fragmentation of a composite nucleus (produced by
the collision of two atomic nuclei). When the nucleus is weakly excited, it
"evaporates" some light particles so that, at the end of the process, we are
left with {\it light fragments} and a heavy {\it evaporation residue} which
contains almost all the charge of the initial nucleus. At very high excitation
energies, the nucleus is completely "vaporized" into light fragments. In the
energy range between these two extremes, the nucleus undergoes
"multifragmentation" into a large number of fragments of all sizes. The
multifragmentation process is thus characterized by the production of {\sc
imf}. Some multifragmentation models consider the light fragments (with charge
less than or equal to 2) as nuclear matter in the {\it gazeous phase} whereas
the evaporation residue or the {\sc imf} form the {\it liquid phase}
\cite{Bib_Col00}. This terminology is also used in the field of phase
transition models. Thus, in various fields, classes of fragments are defined by
their size. Hence it is of interest to enumerate the number of partitions with
constraints on the fragment size.

In Section \ref{Sec sans condition} some techniques used to calculate the number
of partitions of an integer without conditions on the size of the fragments
will be reviewed. It will then be shown how the partitions with constraints on
the maximum size of the fragments (Section \ref{Sec s_max}), on the minimum
size of the fragments (Section \ref{Sec s_min}) and on a size range (Section
\ref{Sec s_min s_max}) can be enumerated.

\subsection {The number of unconstrained partitions}

\label{Sec sans condition}

The total number of partitions of the integer $S$ is given approximatively
by the Ramanujan-Hardy \cite{Bib_Abr65} formula which leading term is :

\begin{equation}
N(S)\approx\frac{\exp{\left(\pi\sqrt{\frac{2S}{3}}\right)}}{4S\sqrt{3}}.
\end{equation}

As can be seen, the number of partitions increases very rapidly with $S$. The
exact value of the number of partitions can be obtained using one of the
following recursive formulae (the number of partitions of $S$ into $M$ fragments
is noted $N(S,M)$ and $M$ is referred to as the multiplicity)~:

\begin{eqnarray}
\label{Eq N(S) rec}
N(S,M) &=& N(S-1,M-1) + N(S-M,M)\\
\label{Eq N(S) rec nouv}
       &=& \sum_{m=1}^M N(S-M,m).
\end{eqnarray}

From this relation \cite{Bib_Bon85}, the equation giving the total number of
partitions can be deduced~:

\begin{equation}
\label{Eq N(S) Bondorf} 
N(S) = 1 + \sum_{M=2}^S \sum_{k=0}^{{\rm Int}(S/M)-1} N(S-kM-1,M-1).
\end{equation}

The Euler \cite{Bib_Rio67_Sob84} recursive relation leads to the same result~:

\begin{eqnarray}
\nonumber
N(S) = \sum_{k=1} (-1)^{k+1}(N(S-\frac{3k^2-k}{2})\\
+N(S-\frac{3k^2+k}{2})).
\label{Eq N(S) Euler} 
\end{eqnarray}

\section{Number of partitions with constraints on the size of the fragments}

\subsection{Constraint on the maximum size}

\label{Sec s_max}

The number of partitions of an integer $S$ into $M$ fragments with size less
than or equal to $s_{max}$ will be noted $^{s_{\rm max}}N(S,M)$. It is obtained
using a modified version of the recursive relation (\ref{Eq N(S) rec})~:

\begin{eqnarray}
\label{Eq N(S_smin_imf) rec}
^{s_{\rm max}}N(S,M) = ^{s_{\rm max}}N(S-1,M-1)\cr
+^{s_{\rm max}}N(S-M,M)\cr
-^{s_{\rm max}}N(S-M-s_{\rm max},M-1)\cr
\hbox{if}\quad S\leq\frac{M\,(s_{\rm max}+1)}{2}, \\
\label{Eq N(S_smin_imf) sym}
^{s_{\rm max}}N(S,M) = ^{s_{\rm max}}N(M\,(s_{\rm max}+1)-S,M).
\end{eqnarray}

The boundary condition is~:

\begin{eqnarray}
^{s_{\rm max}}N(0,1) &=& 1.
\end{eqnarray}

These relations lead to the calculation of $^{s_{\rm max}}N$ for any value of
$S$ and $M$. The three right hand terms in Eq. (\ref{Eq N(S_smin_imf) rec}) are
explained as follows. The partition ensemble can be shared into two sub-groups.
The first one contains all the partitions including at least one size 1
fragment. One of these size 1 fragments can be removed from each partition in
the sub-group. It follows that the number of partitions in the sub-group can be
written as $^{s_{\rm max}}N(S-1,M-1)$. The second group includes the partitions
with no size 1 fragment. Hence one unit can be removed from each fragment
without modifying the multiplicity. In the absence of any condition on the
maximum size, the number of partitions in the second group would be $N(S-M,M)$.
However, among the partitions into $M$ fragments of the integer $S-M$, some
have one or more fragments with size $s_{\rm max}$. It is not possible to add
one unit to these fragments, thus the corresponding partitions should not be
counted. The number of these invalid partitions is obtained by removing one
fragment with size  $s_{\rm max}$. The other fragments can have any size less
than or equal to $s_{\rm max}$ and their multiplicity is $M-1$. The number of
invalid partitions is thus $^{s_{\rm max}}N(S-M-s_{\rm max},M-1)$.

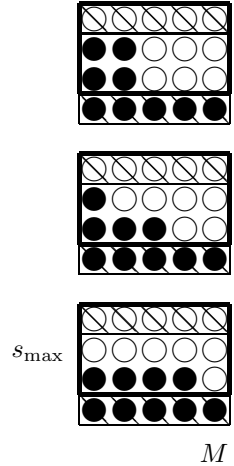
\begin{figure}

\unitlength=1cm

\begin{picture}(2,7.5)(-2,-1)


\put (-.2,0.2){\line(1,-1){.400}}
\put (0.2,0.2){\line(1,-1){.400}}
\put (0.6,0.2){\line(1,-1){.400}}
\put (1.0,0.2){\line(1,-1){.400}}
\put (1.4,0.2){\line(1,-1){.400}}

\put (-.2,1.4){\line(1,-1){.400}}
\put (0.2,1.4){\line(1,-1){.400}}
\put (0.6,1.4){\line(1,-1){.400}}
\put (1.0,1.4){\line(1,-1){.400}}
\put (1.4,1.4){\line(1,-1){.400}}

\put (0.0,0.0){\circle*{.3}}
\put (0.0,0.4){\circle*{.3}}
\put (0.0,0.8){\circle{.3}}
\put (0.0,1.2){\circle{.3}}

\put (0.4,0.0){\circle*{.3}}
\put (0.4,0.4){\circle*{.3}}
\put (0.4,0.8){\circle{.3}}
\put (0.4,1.2){\circle{.3}}

\put (0.8,0.0){\circle*{.3}}
\put (0.8,0.4){\circle*{.3}}
\put (0.8,0.8){\circle{.3}}
\put (0.8,1.2){\circle{.3}}

\put (1.2,0.0){\circle*{.3}}
\put (1.2,0.4){\circle*{.3}}
\put (1.2,0.8){\circle{.3}}
\put (1.2,1.2){\circle{.3}}

\put (1.6,0.0){\circle*{.3}}
\put (1.6,0.4){\circle{.3}}
\put (1.6,0.8){\circle{.3}}
\put (1.6,1.2){\circle{.3}}

\put (-.2,-.2){\line(1,0){2.}}
\put (-.2,-.2){\line(0,1){1.2}}
\put (-.2,1.0){\line(1,0){2.}}
\put (1.8,-.2){\line(0,1){1.2}}

\put (1.4,-.7){$M$}
\put (-1.1,0.7){$s_{\rm max}$}

\thicklines
\linethickness{0.5mm}

\put (-.2,0.2){\line(1,0){2.}}
\put (-.2,0.2){\line(0,1){1.2}}
\put (-.2,1.4){\line(1,0){2.}}
\put (1.8,0.2){\line(0,1){1.2}}


\thinlines

\put (-.2,2.2){\line(1,-1){.400}}
\put (0.2,2.2){\line(1,-1){.400}}
\put (0.6,2.2){\line(1,-1){.400}}
\put (1.0,2.2){\line(1,-1){.400}}
\put (1.4,2.2){\line(1,-1){.400}}

\put (-.2,3.4){\line(1,-1){.400}}
\put (0.2,3.4){\line(1,-1){.400}}
\put (0.6,3.4){\line(1,-1){.400}}
\put (1.0,3.4){\line(1,-1){.400}}
\put (1.4,3.4){\line(1,-1){.400}}

\put (0.0,2.0){\circle*{.3}}
\put (0.0,2.4){\circle*{.3}}
\put (0.0,2.8){\circle*{.3}}
\put (0.0,3.2){\circle{.3}}

\put (0.4,2.0){\circle*{.3}}
\put (0.4,2.4){\circle*{.3}}
\put (0.4,2.8){\circle{.3}}
\put (0.4,3.2){\circle{.3}}

\put (0.8,2.0){\circle*{.3}}
\put (0.8,2.4){\circle*{.3}}
\put (0.8,2.8){\circle{.3}}
\put (0.8,3.2){\circle{.3}}

\put (1.2,2.0){\circle*{.3}}
\put (1.2,2.4){\circle{.3}}
\put (1.2,2.8){\circle{.3}}
\put (1.2,3.2){\circle{.3}}

\put (1.6,2.0){\circle*{.3}}
\put (1.6,2.4){\circle{.3}}
\put (1.6,2.8){\circle{.3}}
\put (1.6,3.2){\circle{.3}}

\put (-.2,1.8){\line(1,0){2.}}
\put (-.2,1.8){\line(0,1){1.2}}
\put (-.2,3.0){\line(1,0){2.}}
\put (1.8,1.8){\line(0,1){1.2}}

\thicklines
\linethickness{0.5mm}

\put (-.2,2.2){\line(1,0){2.}}
\put (-.2,2.2){\line(0,1){1.2}}
\put (-.2,3.4){\line(1,0){2.}}
\put (1.8,2.2){\line(0,1){1.2}}


\thinlines

\put (-.2,4.2){\line(1,-1){.400}}
\put (0.2,4.2){\line(1,-1){.400}}
\put (0.6,4.2){\line(1,-1){.400}}
\put (1.0,4.2){\line(1,-1){.400}}
\put (1.4,4.2){\line(1,-1){.400}}

\put (-.2,5.4){\line(1,-1){.400}}
\put (0.2,5.4){\line(1,-1){.400}}
\put (0.6,5.4){\line(1,-1){.400}}
\put (1.0,5.4){\line(1,-1){.400}}
\put (1.4,5.4){\line(1,-1){.400}}

\put (0.0,4.0){\circle*{.3}}
\put (0.0,4.4){\circle*{.3}}
\put (0.0,4.8){\circle*{.3}}
\put (0.0,5.2){\circle{.3}}

\put (0.4,4.0){\circle*{.3}}
\put (0.4,4.4){\circle*{.3}}
\put (0.4,4.8){\circle*{.3}}
\put (0.4,5.2){\circle{.3}}

\put (0.8,4.0){\circle*{.3}}
\put (0.8,4.4){\circle{.3}}
\put (0.8,4.8){\circle{.3}}
\put (0.8,5.2){\circle{.3}}

\put (1.2,4.0){\circle*{.3}}
\put (1.2,4.4){\circle{.3}}
\put (1.2,4.8){\circle{.3}}
\put (1.2,5.2){\circle{.3}}

\put (1.6,4.0){\circle*{.3}}
\put (1.6,4.4){\circle{.3}}
\put (1.6,4.8){\circle{.3}}
\put (1.6,5.2){\circle{.3}}

\put (-.2,3.8){\line(1,0){2.}}
\put (-.2,3.8){\line(0,1){1.2}}
\put (-.2,5.0){\line(1,0){2.}}
\put (1.8,3.8){\line(0,1){1.2}}

\thicklines
\linethickness{0.5mm}

\put (-.2,4.2){\line(1,0){2.}}
\put (-.2,4.2){\line(0,1){1.2}}
\put (-.2,5.4){\line(1,0){2.}}
\put (1.8,4.2){\line(0,1){1.2}}

\end{picture}
\caption{
\label{Fig boite 1}
Graphical sketchs of the three partitions (\{3,3,1,1,1\}, \{3,2,2,1,1\},
\{2,2,2,2,1\}) of the integer 9 into 5 fragments with size less than or equal
to 3 (thin line boxes). The bold line boxes contain all the complementary
partitions (\{3,3,3,1,1\}, \{3,3,2,2,1\}, \{3,2,2,2,2\}) of the 11 open dots
with the same multiplicity and maximum size. The hatched dots do not 
participate in the enumeration of partitions (due to the multiplicity
constraint).}
\end{figure}

The symmetry relation (\ref{Eq N(S_smin_imf) sym}) can be demonstrated
graphically using the so-called Ferrers diagram in which a size $s$ fragment is
representated by a column of $s$ dots and a partition by its set of fragments
sorted in a decreasing order. We complete the Ferrers diagrams with open dots
as indicated in figure \ref{Fig boite 1}. The number of partitions of the
integer $S$ into $M$ parts with sizes less than or equal to $s_{\rm max}$ is
equal to the number of ways of arranging the dots in the thin line boxe. As the
multiplicity is fixed, the bottom row is necessarily full. By a $180\deg$
rotation the open dots play the same role as the black dots. The open dots
partition will be refered to as {\em complementary} partition of the black dots
(which should not be confused with the conjugate partition which is obtained by
inverting $M$ and $s_{max}$). Thus, to each multiplicity $M$ partition of the
integer $S$ corresponds exactly one multiplicity $M$ partition of the
$M\,(s_{\rm max}+1)-S$ open dots.

For example, in the frame of the minimal information model, it can be
interesting to know the number of partitions containing a given set ${\bf
N}:(N_{s_{\rm min}},\ldots,N_{S})$ of "large" fragments (i.e. fragments with
size greater or equal to $s_{\rm min}$) supplemented by "small" fragments~:

\begin{equation}
N({\bf N})=^{s_{\rm min}-1}N(S-\sum_{s = s_{\rm min}}^S
s\,N_s)\,.
\end{equation}

An alternative method for enumerating the partitions with constraint on the
maximum size consists in using the equivalent of equation (\ref{Eq N(S) rec
nouv}) which takes the following form in this case~:

\begin{eqnarray}
\label{Eq smaxN(S) rec nouv}
^{s_{\rm max}}N(S,M) = \sum_{m=1}^M\ ^{s_{\rm max}-1}N(S-M,m)\,.
\end{eqnarray}

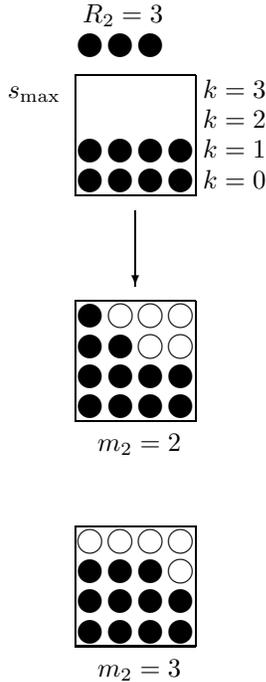
\begin{figure}

\unitlength=1cm
\begin{picture}(2,10.5)(-2,-1)

\thinlines

\put (0.0,6.6){\circle*{.3}}
\put (0.0,7.0){\circle*{.3}}
\put (0.0,8.4){\circle*{.3}}

\put (0.4,6.6){\circle*{.3}}
\put (0.4,7.0){\circle*{.3}}
\put (0.4,8.4){\circle*{.3}}

\put (0.8,6.6){\circle*{.3}}
\put (0.8,7.0){\circle*{.3}}
\put (0.8,8.4){\circle*{.3}}

\put (1.2,6.6){\circle*{.3}}
\put (1.2,7.0){\circle*{.3}}

\put (-.2,6.4){\line(1,0){1.6}}
\put (-.2,6.4){\line(0,1){1.6}}
\put (1.4,6.4){\line(0,1){1.6}}
\put (-.2,8.0){\line(1,0){1.6}}

\put (-1.1,7.7){$s_{\rm max}$}
\put (1.5,6.5){$k=0$}
\put (1.5,6.9){$k=1$}
\put (1.5,7.3){$k=2$}
\put (1.5,7.7){$k=3$}
\put (-.1,8.7){$R_2=3$}

\put (0.6,6.2){\vector(0,-1){1.0}}


\put (0.0,3.6){\circle*{.3}}
\put (0.0,4.0){\circle*{.3}}
\put (0.0,4.4){\circle*{.3}}
\put (0.0,4.8){\circle*{.3}}

\put (0.4,3.6){\circle*{.3}}
\put (0.4,4.0){\circle*{.3}}
\put (0.4,4.4){\circle*{.3}}
\put (0.4,4.8){\circle{.3}}

\put (0.8,3.6){\circle*{.3}}
\put (0.8,4.0){\circle*{.3}}
\put (0.8,4.4){\circle{.3}}
\put (0.8,4.8){\circle{.3}}

\put (1.2,3.6){\circle*{.3}}
\put (1.2,4.0){\circle*{.3}}
\put (1.2,4.4){\circle{.3}}
\put (1.2,4.8){\circle{.3}}

\put (-.2,3.4){\line(1,0){1.6}}
\put (-.2,3.4){\line(0,1){1.6}}
\put (1.4,3.4){\line(0,1){1.6}}
\put (-.2,5.0){\line(1,0){1.6}}

\put (0.1,3.0){$m_2=2$}


\put (0.0,0.6){\circle*{.3}}
\put (0.0,1.0){\circle*{.3}}
\put (0.0,1.4){\circle*{.3}}
\put (0.0,1.8){\circle{.3}}

\put (0.4,0.6){\circle*{.3}}
\put (0.4,1.0){\circle*{.3}}
\put (0.4,1.4){\circle*{.3}}
\put (0.4,1.8){\circle{.3}}

\put (0.8,0.6){\circle*{.3}}
\put (0.8,1.0){\circle*{.3}}
\put (0.8,1.4){\circle*{.3}}
\put (0.8,1.8){\circle{.3}}

\put (1.2,0.6){\circle*{.3}}
\put (1.2,1.0){\circle*{.3}}
\put (1.2,1.4){\circle{.3}}
\put (1.2,1.8){\circle{.3}}

\put (-.2,0.4){\line(1,0){1.6}}
\put (-.2,0.4){\line(0,1){1.6}}
\put (1.4,0.4){\line(0,1){1.6}}
\put (-.2,2.0){\line(1,0){1.6}}

\put (0.1,0.0){$m_2=3$}

\end{picture}
\caption{
\label{Fig mk}
Ferrers diagrams illustrating the determination of the minima and maxima of the
sums in equation (\ref{smaxN(S) sum}). The goal is to determine the range of
$m_2$ when $m_0=m_1=4$ and $S=11$ (thus $R_2=3$). The fragments are sorted in
decreasing order, thus $m_2\,(s_{\rm max}-2)\geq R_2$, that is, the minimum
value of $m_2$ is 2. The number of fragments with size greater than $k$ is
necessarily lower than the number of fragments with size greater than $k-1$,
thus $m_2\leq m_1$. Furthermore, their are only $R_2$ units left, thus $m_2\leq
R_2$. Finally, the maximum value for $m_2$ is the minimum of $m_1$ and $R_2$
(i.e. 3). More generally, the sum for $m_k$ runs from Int($\frac{R_k-1}{s_{\rm
max}-k})+1$ to Min($m_{k-1},R_k$).}

\end{figure}

This equation can be applied recursively $s_{\rm max}-1$ times so that the
maximum size in the right hand term is 1. Using $^1N(S,M)=1$ if $S=M$ and 0
otherwise, one obtains~:

\begin{eqnarray}
\label{smaxN(S) sum}
^{s_{\rm max}}N(S,M) = 
\sum_{m_1} \ldots 
\sum_{m_k = {\rm Int}(\frac{R_k-1}{s_{\rm max}-k})+1}^{{\rm Min}(m_{k-1},R_k)} 
\ldots 
\sum_{m_{s_{\rm max}-2}} 1\,,
\end{eqnarray}

with $m_0=M$ and $R_k=S-\sum_{i=0}^{k-1} m_i$. In this equation, $m_k$ is the
multiplicity of fragments with size strictly greater than $k$. The determination
of the range for $m_k$ is illustrated in Fig. \ref{Fig mk}.

Using the same line of thought on the conjugate partition, one obtains the
following equation~:

\begin{equation}
\label{smaxN(S) sum'}
^{s_{\rm max}}N(S,M) = 
\sum_{s_1} \ldots 
\sum_{s_k = {\rm Int}(\frac{R_k-1}{M-k+1})+1}^{{\rm Min}(s_{k-1},R_k-M+1)} 
\ldots 
\sum_{s_{M-1}} 1\,,
\end{equation}

with $s_0 = s_{\rm max}$ and $R_k = S -\sum_{i=1}^{k-1}s_i$, $s_i$ being the
size of the $i^{\rm th}$ largest fragment. The same equations hold for $N(S,M)$,
fixing $s_{\rm max} = S$.

\subsection{Constraint on the minimum size}

\label{Sec s_min}

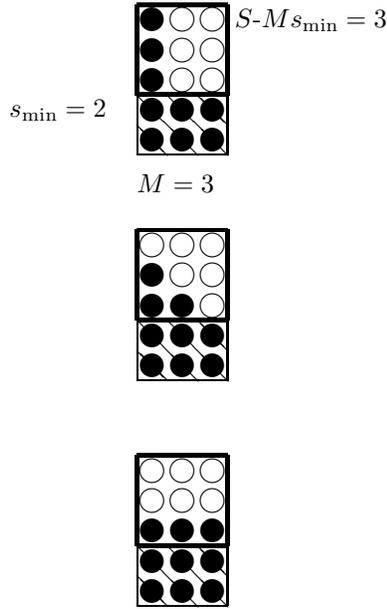
\begin{figure}

\unitlength=1cm

\begin{picture}(2,8.5)(-2,-1)


\thinlines

\put (-.2,6.2){\line(1,-1){.40}}
\put (-.2,6.6){\line(1,-1){.81}}
\put (0.2,6.6){\line(1,-1){.81}}
\put (0.6,6.6){\line(1,-1){.40}}

\put (0.0,6.0){\circle*{.3}}
\put (0.0,6.4){\circle*{.3}}
\put (0.0,6.8){\circle*{.3}}
\put (0.0,7.2){\circle*{.3}}
\put (0.0,7.6){\circle*{.3}}

\put (0.4,6.0){\circle*{.3}}
\put (0.4,6.4){\circle*{.3}}
\put (0.4,6.8){\circle{.3}}
\put (0.4,7.2){\circle{.3}}
\put (0.4,7.6){\circle{.3}}

\put (0.8,6.0){\circle*{.3}}
\put (0.8,6.4){\circle*{.3}}
\put (0.8,6.8){\circle{.3}}
\put (0.8,7.2){\circle{.3}}
\put (0.8,7.6){\circle{.3}}

\put (-.2,5.8){\line(1,0){1.2}}
\put (-.2,5.8){\line(0,1){2.}}
\put (1.0,5.8){\line(0,1){2.}}

\thicklines
\linethickness{0.5mm}

\put (-.2,6.6){\line(1,0){1.2}}
\put (-.2,6.6){\line(0,1){1.2}}
\put (-.2,7.8){\line(1,0){1.2}}
\put (1.0,6.6){\line(0,1){1.2}}

\put (-.2,5.3){$M=3$}
\put (-1.9,6.3){$s_{\rm min}=2$}
\put (1.1,7.5){$S$-$Ms_{\rm min}=3$}


\thinlines

\put (-.2,3.2){\line(1,-1){.40}}
\put (-.2,3.6){\line(1,-1){.81}}
\put (0.2,3.6){\line(1,-1){.81}}
\put (0.6,3.6){\line(1,-1){.40}}

\put (0.0,3.0){\circle*{.3}}
\put (0.0,3.4){\circle*{.3}}
\put (0.0,3.8){\circle*{.3}}
\put (0.0,4.2){\circle*{.3}}
\put (0.0,4.6){\circle{.3}}

\put (0.4,3.0){\circle*{.3}}
\put (0.4,3.4){\circle*{.3}}
\put (0.4,3.8){\circle*{.3}}
\put (0.4,4.2){\circle{.3}}
\put (0.4,4.6){\circle{.3}}

\put (0.8,3.0){\circle*{.3}}
\put (0.8,3.4){\circle*{.3}}
\put (0.8,3.8){\circle{.3}}
\put (0.8,4.2){\circle{.3}}
\put (0.8,4.6){\circle{.3}}

\put (-.2,2.8){\line(1,0){1.2}}
\put (-.2,2.8){\line(0,1){2.}}
\put (1.0,2.8){\line(0,1){2.}}

\thicklines
\linethickness{0.5mm}

\put (-.2,3.6){\line(1,0){1.2}}
\put (-.2,3.6){\line(0,1){1.2}}
\put (-.2,4.8){\line(1,0){1.2}}
\put (1.0,3.6){\line(0,1){1.2}}


\thinlines

\put (-.2,0.2){\line(1,-1){.40}}
\put (-.2,0.6){\line(1,-1){.81}}
\put (0.2,0.6){\line(1,-1){.81}}
\put (0.6,0.6){\line(1,-1){.40}}

\put (0.0,0.0){\circle*{.3}}
\put (0.0,0.4){\circle*{.3}}
\put (0.0,0.8){\circle*{.3}}
\put (0.0,1.2){\circle{.3}}
\put (0.0,1.6){\circle{.3}}

\put (0.4,0.0){\circle*{.3}}
\put (0.4,0.4){\circle*{.3}}
\put (0.4,0.8){\circle*{.3}}
\put (0.4,1.2){\circle{.3}}
\put (0.4,1.6){\circle{.3}}

\put (0.8,0.0){\circle*{.3}}
\put (0.8,0.4){\circle*{.3}}
\put (0.8,0.8){\circle*{.3}}
\put (0.8,1.2){\circle{.3}}
\put (0.8,1.6){\circle{.3}}

\put (-.2,-.2){\line(1,0){1.2}}
\put (-.2,-.2){\line(0,1){2.}}
\put (1.0,-.2){\line(0,1){2.}}

\thicklines
\linethickness{0.5mm}

\put (-.2,0.6){\line(1,0){1.2}}
\put (-.2,0.6){\line(0,1){1.2}}
\put (-.2,1.8){\line(1,0){1.2}}
\put (1.0,0.6){\line(0,1){1.2}}

\end{picture}
\caption{
\label{Fig boite 2}
Graphical sketchs of all the partitions (\{5,2,2\}, \{4,3,2\}, \{3,3,3\}) of
the integer 9 into 3 fragments with size greater or equal to 2. The two lower
rows play no role in the counting of the partitions. The bold line box includes
all  the partitions of the integer 3 (i.e. $S-M\,s_{\rm min}$) with
multiplicities less than or equal to 3.}
\end{figure}

The number of partitions of the integer $S$ into $M$ fragments with size
greater or equal to $s_{\rm min}$ will be noted $_{s_{\rm min}}N(S,M)$. In each
event, $M\,s_{\rm min}$ units are imposed (in Fig. \ref{Fig boite 2} they
correspond to the two lower rows). The number of partitions only depends on the
$S-M\,s_{\rm min}$ remaining units, for multiplicities ranging from 1 to $M$~:

\begin{equation}
_{s_{\rm min}}N(S,M) = \sum_{m=1}^M N(S-M\,s_{\rm min},m).
\end{equation}

Following Eq. (\ref{Eq N(S) rec nouv}), this expression can be simplified
to~:

\begin{equation}
_{s_{\rm min}}N(S,M) = N(S-M(s_{\rm min}-1),M).
\end{equation}

The same property can be directly deduced by considering the complementary
partition (see open dots in Fig. \ref{Fig boite 2}).

The total number of partitions is~:

\begin{equation}
\label{Eq N(S) smin}
\begin{array}{lll}
_{s_{\rm min}}N(S) &=& 
\sum_{M=1}^{S/s_{\rm min}}\ N(S-M(s_{\rm min}-1),M).
\end{array}
\end{equation}

The boundary conditions are~:

\begin{equation}
\begin{array}{lll}
N(0,M \ne 1) &=& 0 \quad \hbox{and}\cr
N(0,1)       &=& 1.
\end{array}
\end{equation}

\subsection{Constraint on the minimum and maximum sizes}

\label{Sec s_min s_max}

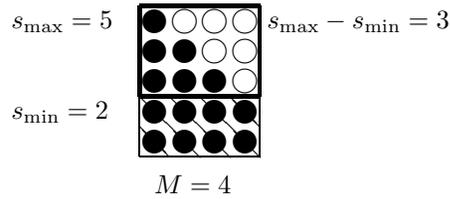
\begin{figure}

\unitlength=1cm

\begin{picture}(2,3)(-2,-1)

\thinlines

\put (-.2,0.2){\line(1,-1){.400}}
\put (-.2,0.6){\line(1,-1){.81}}
\put (0.2,0.6){\line(1,-1){.81}}
\put (0.6,0.6){\line(1,-1){.81}}
\put (1.0,0.6){\line(1,-1){.400}}

\put (0.0,0.0){\circle*{.3}}
\put (0.0,0.4){\circle*{.3}}
\put (0.0,0.8){\circle*{.3}}
\put (0.0,1.2){\circle*{.3}}
\put (0.0,1.6){\circle*{.3}}

\put (0.4,0.0){\circle*{.3}}
\put (0.4,0.4){\circle*{.3}}
\put (0.4,0.8){\circle*{.3}}
\put (0.4,1.2){\circle*{.3}}
\put (0.4,1.6){\circle{.3}}

\put (0.8,0.0){\circle*{.3}}
\put (0.8,0.4){\circle*{.3}}
\put (0.8,0.8){\circle*{.3}}
\put (0.8,1.2){\circle{.3}}
\put (0.8,1.6){\circle{.3}}

\put (1.2,0.0){\circle*{.3}}
\put (1.2,0.4){\circle*{.3}}
\put (1.2,0.8){\circle{.3}}
\put (1.2,1.2){\circle{.3}}
\put (1.2,1.6){\circle{.3}}

\put (-.2,-.2){\line(1,0){1.6}}
\put (-.2,-.2){\line(0,1){2.0}}
\put (-.2,1.8){\line(1,0){1.6}}
\put (1.4,-.2){\line(0,1){2.0}}

\thicklines
\linethickness{0.5mm}

\put (-.2,0.6){\line(1,0){1.6}}
\put (-.2,0.6){\line(0,1){1.2}}
\put (-.2,1.8){\line(1,0){1.6}}
\put (1.4,0.6){\line(0,1){1.2}}

\put (0.0,-.7){$M=4$}
\put (-1.9,0.3){$s_{\rm min}=2$}
\put (-1.9,1.5){$s_{\rm max}=5$}
\put (1.5,1.5){$s_{\rm max}-s_{\rm min}=3$}

\end{picture}
\caption{
\label{Fig boite 3}
Graphical sketch of one partition (\{5,4,3,2\}) of the integer 14 into fragments
with size included between 2 and 5. The two lower rows play no role in the
partition counting. The bold line box contains all the partitions of the integer
6 (i.e. $S-M\,s_{\rm min}$) into fragments with size less than or equal to 3 and 
multiplicity less than or equal to 4.}
\end{figure}

When both the minimum and the maximum size of the fragments are fixed, the
counting of the partitions is carried out in the same way as previously :
$M\,s_{\rm min}$ units play no role. The number of partitions is the same as
that of the integer $S-M\,s_{\rm min}$ into fragments with size less than or
equal to $s_{\rm max}-s_{\rm min}$ (Fig. \ref{Fig boite 3}). Thus, the number
of doubly-conditioned partitions is obtained as a sum over the number of
singly-conditioned partitions.

\begin{eqnarray}
\label{Eq N(S) smin smax 1}
_{s_{\rm min}}^{s_{\rm max}}N(S,M) &=&
\sum_{m=1}^M {^{s_{\rm max}-s_{\rm min}}}N(S-M\,s_{\rm min},m).
\end{eqnarray}

An alternative expression can be obtained by considering the complementary
partitions (see Fig. \ref{Fig boite 3})~:

\begin{eqnarray}
\label{Eq N(S) smin smax 2}
_{s_{\rm min}}^{s_{\rm max}}N(S,M) &=&
\sum_{m=1}^M {^{s_{\rm max}-s_{\rm min}}}N(M\,s_{\rm max}-S,m).
\end{eqnarray}

Applying Eq. (\ref{Eq smaxN(S) rec nouv}) to (\ref{Eq N(S) smin smax 1}) and
(\ref{Eq N(S) smin smax 2}) one obtains respectively~:

\begin{eqnarray}
\label{Eq N(S) smin smax 3}
_{s_{\rm min}}^{s_{\rm max}}N(S,M)
&=&{^{s_{\rm max}-s_{\rm min}+1}}N(S-M(s_{\rm min}+1),M)\,,\\
\label{Eq N(S) smin smax 4}
&=&{^{s_{\rm max}-s_{\rm min}+1}}N(M(s_{\rm max}+1)-S,M)\,.
\end{eqnarray}

Finally~:

\begin{eqnarray}
_{s_{\rm min}}^{s_{\rm max}}N(S) &=&
\sum_{M=1}^{S/s_{\rm min}} 
{_{s_{\rm min}}^{s_{\rm max}}}N(S,M).
\end{eqnarray}

\section {Conclusion}

In this article, we have provided formulae for the calculation of the number of
partitions with conditions on the maximum fragment size (Eq. (\ref{Eq
N(S_smin_imf) rec},\ref{Eq N(S_smin_imf) sym})), with conditions on the minimum
fragment size (Eq. (\ref{Eq N(S) smin})) and with conditions on both the
minimum and the maximum fragment size (Eq. (\ref{Eq N(S) smin smax 3},\ref{Eq
N(S) smin smax 4})). To demonstrate these formulae, the notion of complementary
partitions was introduced. The constrained partition numbers are notably useful
in the analysis of nuclear multifragmentation. Moretto and collaborators
\cite{Bib_Mor96_Mor97} have introduced an elegant combinatorial procedure to
isolate rare events corresponding to the fragmentation of the atomic nucleus in
a number of nearly equal size {\sc imf} (fragments with charge greater or equal
to $Z_{\rm min}$) supplemented by light fragments (fragments with charge less
than or equal to $Z_{\rm min}-1$). This procedure requires the evaluation of
the number of partitions corresponding to a given sum ($Z_{\rm imf}$) of the
charges of a given number ($M$) of {\sc imf}. This number of partitions is
given as $_{Z_{\rm min}}N(Z_{\rm imf},M)\ ^{Z_{\rm min}-1}N(Z_{\rm tot}-Z_{\rm
imf})$. The total number of partitions can be evaluated by the following
convolution $N(S) = \sum_{s}\,_{s_{\rm min}}N(s)\ ^{s_{\rm min}-1}N(S-s)$. In a
forthcomming article \cite{Bib_Des01} we will show how the Moretto charge
correlation can be calculated explicitly in the frame of the minimal
information model. More generally these formulae are useful in domains where
the fragment classes ({\it infinite fragments, evaporation residues, light
particles, intermediate mass fragments, liquid} and {\it gazeous phases}
\ldots) are defined with respect to their sizes.

\end{document}